\documentclass[intlimits,twoside,a4paper]{article}

\usepackage{amsmath,amssymb}
\usepackage{graphicx}
\usepackage{wrapfig}

\usepackage[T2A]{fontenc}
\usepackage[cp1251]{inputenc}
\usepackage{color}
\usepackage[eqsecnum]{cmpj2}

\issue{2013}{16}{1}{13601}
\doinumber{10.5488/CMP.16.13601}

\title[$XYZ$-Heisenberg-Ising two-leg ladder]%
{Ground state of a spin-1/2 Heisenberg-Ising two-leg ladder with $XYZ$ intra-rung coupling}
\author[T.~Verkholyak, J.~Stre\v{c}ka]{T.~Verkholyak\refaddr{label1},
        J.~Stre\v{c}ka\refaddr{label2}}
\addresses{
\addr{label1} Institute for Condensed Matter Physics
of the National Academy of Sciences of Ukraine,\\
1 Svientsitskii St., 79011 Lviv, Ukraine
\addr{label2} Department of Theoretical Physics and Astrophysics,
             Faculty of Science, P. J. \v{S}af\'{a}rik University,\\
             Park Angelinum 9, 040 01 Ko\v{s}ice, Slovak Republic
}

\date{Received July 3, 2012}

\begin{document}

\maketitle

\begin{abstract}
The quantum spin-1/2 two-leg ladder with an anisotropic $XYZ$ Heisenberg intra-rung interaction
and Ising inter-rung interactions is treated by means of a rigorous approach based on the unitary
transformation. The particular case of the considered model with $X$--$X$ intra-rung interaction
resembles a quantum compass ladder with additional frustrating diagonal Ising interactions.
Using an appropriately chosen unitary transformation, the model under investigation may be reduced
to a transverse Ising chain with composite spins, and one may subsequently find the ground state
quite rigorously. We obtain a ground-state phase diagram and analyze the interplay of the competition
between several factors: the $XYZ$ anisotropy in the Heisenberg intra-rung coupling,
the Ising interaction along the legs, and the frustrating diagonal Ising interaction. The investigated
model shows extraordinarily diverse ground-state phase diagrams including several unusual quantum
ordered phases, two different disordered quantum paramagnetic phases, as well as discontinuous or continuous quantum
phase transitions between those phases.
\pacs 05.30.Rf, 75.10.Jm
\end{abstract}

\section{Introduction}
Quantum spin ladders with frustrated interactions are intensively studied during the last few decades, since they exhibit a rather complex ground-state behaviour to be reflected in extraordinarily rich low-temperature thermodynamics as well \cite{solstsci164,lnp645}. Quite recently, a number of exact solutions have been obtained for several particular examples of quantum spin-$\frac{1}{2}$ two-leg ladders \cite{brze2009,brze10a,brze10b,lai2011,verkholyak2012}. The railroad ladder considered by Lai and Montrunich \cite{lai2011} has a quite specific configuration of inter-spin interactions, namely, the staggering of $X$--$X$ and $Y$--$Y$ couplings along the legs is supplemented by the uniform $Z$--$Z$ coupling present along the rungs. An exact solution of this specific quantum spin ladder has been found by adopting the method originally developed by Kitaev \cite{kitaev20006}, which proved a striking spin-liquid ground state in this quantum spin ladder. On the other hand, the railroad ladder with the uniform $Z$--$Z$ interaction along the legs and the uniform $X$--$X$ interaction along the rungs has been rigorously solved by Brzezicki and Ole\'{s} \cite{brze2009,brze10a,brze10b}. To a certain extent, this exactly solved quantum spin ladder can be regarded as an one-dimensional analogue of the quantum compass model on a square lattice, which describes the orbital ordering in transition-metal compounds \cite{khomskii2003}.

In this work, we will examine a more general model of the quantum spin-$\frac{1}{2}$ two-leg ladder, which includes the fully anisotropic $XYZ$-Heisenberg coupling between spins from the same rung and two different Ising ($Z$--$Z$) interactions between spins from neighbouring rungs considered along the legs and across the diagonals, respectively. Nevertheless, it should be  mentioned that the investigated quantum spin ladder extends our previous exact calculations for the spin-$\frac{1}{2}$ Heisenberg-Ising ladder with the $XXZ$ intra-rung coupling \cite{verkholyak2012}, whereas the quantum compass ladder investigated in detail by Brzezicki and Ole\'{s} \cite{brze2009,brze10a,brze10b} also represents a very special
limiting case of the investigated model system. The main goal of the present paper is to examine the simultaneous effect of two different kinds of frustration: the geometric frustration caused by the antiferromagnetic interaction between spins from an elementary triangle plaquette and the competition between $X$--$X$ and $Y$--$Y$ intra-rung interactions with both $Z$--$Z$ inter-rung interactions.

The outline of the paper is as follows. In section~2, we define the model and show how to get the ground state by a rigorous calculation based on the appropriate unitary transformation. The ground-state phase diagram of the spin-$\frac{1}{2}$ $XZ$-Ising and  $XY$-Ising ladders is explored in section~3. Finally, some conclusions are drawn in section~4.

\section{Model and solution}

\begin{figure}[ht]
\centerline{
\includegraphics[width=0.8\textwidth]{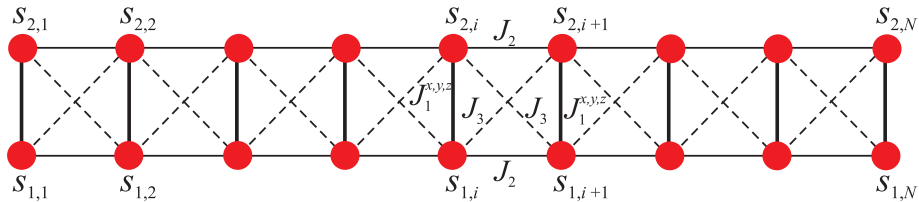}
}
\caption{(Color online) A schematic representation of the considered Heisenberg-Ising two-leg ladder. The intra-rung $XYZ$-Heisenberg coupling is represented by thick lines, while $Z$--$Z$ Ising interactions along the legs and across the diagonals are displayed by thin solid and broken lines, respectively.}
\label{fig1}
\end{figure}

Consider the quantum spin-$\frac{1}{2}$ Heisenberg-Ising ladder with an anisotropic $XYZ$ intra-rung coupling
and two different Ising-type couplings, which involve $Z$--$Z$ spin-spin interactions along the legs
and across the diagonals of a two-leg ladder (see figure~\ref{fig1}):
\begin{equation}
H=\sum_{i=1}^N\left[\left(J_1^x s_{1,i}^x s_{2,i}^x +J_1^y s_{1,i}^y s_{2,i}^y + J_1^z s_{1,i}^z s_{2,i}^z\right)
+J_2\left(s_{1,i}^z s_{1,i+1}^z + s_{2,i}^z s_{2,i+1}^z\right)
+J_3\left(s_{1,i}^z s_{2,i+1}^z + s_{2,i}^z s_{1,i+1}^z\right)
\right].
\label{ham}
\end{equation}
Here, $s_{j,i}^{\alpha}$ denote three spatial components $\alpha=x,y,z$ of the spin-$\frac{1}{2}$ operator,
the former subscript $j=1,2$ determines the number of a leg and the latter subscript enumerates the lattice position
in a particular leg. Apparently, the interaction terms $J_1^x, J_1^y, J_1^z$ account
for the quite anisotropic $XYZ$-Heisenberg coupling between two spins belonging to the same rung,
while the interaction terms $J_2$ and $J_3$ take into consideration the Ising-type interactions
between the nearest-neighbor spins along the legs and across the diagonals of the two-leg ladder.
It should be pointed out that the $z$-component of the total spin $S_i^z=s_{1,i}^z+s_{2,i}^z$
on $i$th rung commutes with the Hamiltonian, i.e. $[(S_i^z)^2,H]=0$, and this property allows
us to present the model in an integrable form.
To obtain a ground state of the investigated quantum spin model quite rigorously,
it is advisable to closely follow the procedure worked out previously
for the quantum spin-$\frac{1}{2}$ Heisenberg-Ising ladder with $XXZ$ intra-rung interaction.
After the unitary transformation \cite{verkholyak2012}
\begin{align}
&{s}_{1,i}^x ={\tilde s}_{1,i}^x\,,
&&{s}_{1,i}^y =2{\tilde s}_{1,i}^y {\tilde s}_{2,i}^x\,,
&&{s}_{1,i}^z =2{\tilde s}_{1,i}^z {\tilde s}_{2,i}^x\,,
\nonumber\\
&{s}_{2,i}^x =2{\tilde s}_{1,i}^x {\tilde s}_{2,i}^z\,,
&&{s}_{2,i}^y =-2{\tilde s}_{1,i}^x {\tilde s}_{2,i}^y\,,
&&{s}_{2,i}^z = {\tilde s}_{2,i}^x\,,
\label{spin-transf}
\end{align}
one may rewrite the Hamiltonian (\ref{ham}) into the following pseudospin representation:
\begin{equation}
H=\sum_{i=1}^N\left\{
\left(\frac{J_1^x}{2}-J_1^y {\tilde s}_{1,i}^z \right){\tilde s}_{2,i}^z
+ \frac{J_1^z}{2} {\tilde s}_{1,i}^z
+\left[J_2\left(1+4{\tilde s}_{1,i}^z {\tilde s}_{1,i+1}^z\right)
+2J_3\left({\tilde s}_{1,i}^z+{\tilde s}_{1,i+1}^z\right)   \right]{\tilde s}_{2,i}^x {\tilde s}_{2,i+1}^x
\right\},
\label{ham2}
\end{equation}
which shows the symmetry of the model in a more explicit way.
It is quite obvious that only $z$-components of spin operators from the first leg are present
in the Hamiltonian (\ref{ham2}), which means that ${\tilde s}_{1,i}^z$ are good quantum numbers.
By contrast, different spatial components of spin operators from the second leg are still involved
in the Hamiltonian (\ref{ham2}) and thus, they still represent quantum spins with regard
to the presence of two non-commuting parts ${\tilde s}_{2,i}^{x}$ and ${\tilde s}_{2,i}^{z}$ of each spin operator.
Altogether, the Hamiltonian (\ref{ham2}) can be identified as the Ising chain of composite spins in a transverse field,
whereas the values of the effective interaction and the effective transverse field locally depend on
a particular choice of eigenvalues of the classical Ising spins ${\tilde s}_{1,i}^z$. Following \cite{verkholyak2012},
one may also establish the following correspondence between new and initial states:
\begin{eqnarray}
&&|{\tilde\downarrow}_{1,i}{\tilde\downarrow}_{2,i}\rangle  = |\phi_{0,0}^{i}\rangle=\frac{1}{\sqrt2}
\left(| \downarrow_{1,i}\uparrow_{2,i}\rangle - | \uparrow_{1,i}\downarrow_{2,i}\rangle\right),\qquad
| {\tilde\downarrow}_{1,i}{\tilde\uparrow}_{2,i}\rangle=|\phi_{1,0}^{i}\rangle=\frac{1}{\sqrt2}
\left(| \downarrow_{1,i}\uparrow_{2,i}\rangle + | \uparrow_{1,i}\downarrow_{2,i}\rangle\right),
\nonumber\\
&&|{\tilde\uparrow}_{1,i}{\tilde\downarrow}_{2,i}\rangle = |\phi_{1,-}^{i}\rangle=\frac{1}{\sqrt2}
\left(| \uparrow_{1,i}\uparrow_{2,i}\rangle - | \downarrow_{1,i}\downarrow_{2,i}\rangle\right),\qquad
|{\tilde\uparrow}_{1,i}{\tilde\uparrow}_{2,i}\rangle=|\phi_{1,+}^{i}\rangle=\frac{1}{\sqrt2}
\left(| \uparrow_{1,i}\uparrow_{2,i}\rangle + | \downarrow_{1,i}\downarrow_{2,i}\rangle\right),
\nonumber\\
\end{eqnarray}
where $|\phi_{0,0}^{i}\rangle$ denotes the singlet state of the $i$th rung and the other three states $|\phi_{1,\pm}^{i}\rangle$,
$|\phi_{1,0}^{i}\rangle$ form the triplet state of the $i$th rung. It should be stressed that the square of $z$th component
of the total spin $S_i^z$ on the $i$th rung acquires two different values. It either equals $(S_i^z)^2=0$ for $|\phi_{0,0}^{i}\rangle$ and $|\phi_{1,0}^{i}\rangle$,
or $(S_i^z)^2=1$ for $|\phi_{1,\pm}^{i}\rangle$.
To get the partition function one has to diagonalize the Hamiltonian (\ref{ham2}) for all particular configurations of ${\tilde s}_{1,i}^z$ and sum up all contributions in the trace of statistical operator. However, it is quite evident from the transformed Hamiltonian (\ref{ham2}) that the chain decomposes into two independent parts whenever two neighboring spins ${\tilde s}_{1,i}^z$ and ${\tilde s}_{1,i+1}^z$ have opposite orientation (i.e., take on different eigenvalues). In this respect, the composite chain is divided into a set of finite chains of different sizes for any chosen configuration of ${\tilde s}_{1,i}^z$. Generally, this problem seems to represent a quite formidable task, but the ground state of the investigated model can be found quite rigorously using the same arguments as given in \cite{verkholyak2012}. Since the ground-state energy of two finite but isolated spin-$\frac{1}{2}$ Ising chains in a transverse field is always higher than the ground-state energy of one unique spin-$\frac{1}{2}$ Ising chain in a transverse field obtained by joining both independent finite chains, the ground state of the model under investigation should accordingly correspond only to the uniform configuration of all ${\tilde s}_{1,i}^z$. Therefore, one may single out only two different uniform configurations with all $\tilde{s}_{1,i}^z=\frac{1}{2}$ or all $\tilde{s}_{1,i}^z=-\frac{1}{2}$ from which the ground state of the Heisenber-Ising ladder can be derived. The effective Hamiltonian (\ref{ham2}) for the two uniform configurations acquires the following form:
\begin{eqnarray}
H^{\pm}=\sum_{i=1}^N \left[
\frac{1}{2}\left(J_1^x\mp J_1^y\right)\tilde{s}_{2,i}^z \pm \frac{1}{4}J_1^z +2\left(J_2\pm J_3\right)\tilde{s}_{2,i}^x \tilde{s}_{2,i+1}^x
\right].
\label{ham-pm}
\end{eqnarray}
The ground state energies per site of both effective Hamiltonians can be  exactly calculated using the Jordan-Wigner fermionization \cite{lsm,pfeuty1970}:
\begin{eqnarray}
e^{\pm}_0==-\frac{\left(J^{\mp}_1+|J_2\pm J_3|\right)}{\pi}{\mathbf E}\left[\sqrt{1-\left(\gamma^{\pm}\right)^2}\right] \pm \frac{J^z_1}{4}\,,
\end{eqnarray}
where
\[\gamma^{\pm}=\frac{J^{\mp}_1-|J_2\pm J_3|}{J^{\mp}_1+|J_2\pm J_3|},\qquad
J^{\pm}=\frac{\left(J^x_1\pm J^y_1\right)}{2}
\]
and
\[
{\mathbf E}(\kappa) = \displaystyle \int_0^{\frac{\pi}{2}} \!\! {\rm d} \theta \sqrt{1 - \kappa^2 \sin^2 \theta}
 \]
 is the complete elliptic integral of the second kind.

Both Hamiltonians $H^+$ and $H^-$
imply a precise mapping correspondence between the spin-$\frac{1}{2}$ Heisen\-berg-Ising two-leg ladder
and the spin-$\frac{1}{2}$ quantum Ising chain in a transverse field,
which can be, however, characterized by different values of the effective interaction and transverse field.
Bearing this in mind, one should expect quantum phase transitions
of two different types. The first kind of zero-temperature phase
transitions may correspond to a continuous (second-order) quantum
phase transition inherent to the transverse Ising chain, which
arises for one particular ratio between the effective interaction
and transverse field. Beside this, there  may also occur
discontinuous (first-order) quantum phase transitions whenever a
crossing of the lowest-energy levels inherent to both effective
Hamiltonians (\ref{ham-pm}) takes place. In the
following two sections, we will illustrate all the aforementioned
features of quantum phase transitions on ground-state phase
diagrams of two particular cases of the model under consideration.

\section{Ground state of $XZ$-Ising and $XY$-Ising ladders}
In this section, we will consider two particular cases of the investigated model (\ref{ham}) by switching off
either the $y$- or $z$-component of $XYZ$-Heisenberg coupling (i.e., either $J_1^y=0$ or $J_1^z=0$).
It is noteworthy that the two aforementioned particular cases represent a direct extension
of the quantum compass ladder \cite{brze2009,brze10a,brze10b} to which the considered model reduces
when neglecting the $z$-component of the Heisenberg coupling ($J_1^z=0$), one of the two transverse
components of the Heisenberg coupling (i.e., either $J_1^x=0$ or $J_1^y=0$) and the frustrating Ising
interaction across the diagonals ($J_3 = 0$). Furthermore, the problem of two-dimensional
quantum compass model is quite complex and the exact solution for this model has not been found yet.

Let us first consider all possible phases that may appear in the ground state of the model under investigation.
Each uniform configuration of $\tilde{s}_{1,i}^z$ corresponds to the transverse Ising chain, which has three possible ground-state phases.
The ground-state phases for all $\tilde{s}_{1,i}^z=-\frac{1}{2}$ appear if $e_0^-<e_0^+$. It is worthwhile to remark that the ground-state phases belonging to this effective model were thoroughly analyzed in our preceding paper \cite{verkholyak2012}
and let us, therefore, give here just their definition for the sake of easy reference:
\begin{itemize}
\item {\bf Quantum paramagnetic (QPM1) state} for $\frac{1}{2}(J_1^x+J_1^y)>|J_2-J_3|$:
    the Heisenberg-Ising ladder is in the gapped disordered state
    with no spontaneous magnetization. The rung singlet dimers
    dominate on the Heisenberg bonds and in the very special case $J_2=J_3$,
    the ground state factorizes to a set of the completely non-correlated rung singlet dimers.
 \item {\bf Stripe Leg (SL) state} for $\frac{1}{2}(J_1^x+J_1^y)<J_3-J_2$:
    the Heisenberg-Ising ladder shows a ferromagnetic order along the legs
    and antiferromagnetic order along the rungs, i.e., the magnetizations of two chains are oriented
    opposite to each other
    ($\langle s_{1,i}^z\rangle=\langle s_{1,i+1}^z\rangle
     =-\langle s_{2,i}^z\rangle=-\langle s_{2,i+1}^z\rangle\neq 0$).
    The following staggered magnetization can be defined as the relevant
    order parameter of this phase
    \[
    m_{\mathrm{SL}}^z=\frac{1}{2N}\sum_{i=1}^N\left(\langle s_{1,i}^z\rangle-\langle s_{2,i}^z\rangle\right)
     =\frac{1}{2}\left[1-\frac{\left(J_1^x+J_1^y\right)^2}{4\left(J_2-J_3\right)^2}\right]^{\frac{1}{8}},
     \]
     which undergoes the obvious quantum reduction of magnetization.
 \item {\bf N\'{e}el state} for $\frac{1}{2}(J_1^x+J_1^y)<J_2-J_3$:
    the nearest-neighbor spins both along the legs and rungs exhibit
    predominantly antiferromagnetic ordering. The dependence of staggered magnetization as the
    relevant order parameter is quite analogous to the previous case
    \[
    m_{\mathrm{AF}}^z=\frac{1}{2N}\sum_{i=1}^N\left(-1\right)^{i}\left(\langle s_{1,i}^z\rangle-\langle s_{2,i}^z\rangle\right)
     =\frac{1}{2}\left[1-\frac{\left(J_1^x+J_1^y\right)^2}{4\left(J_2-J_3\right)^2}\right]^{\frac{1}{8}}.
     \]
\end{itemize}
The most fundamental difference between the ground states of the Heisenberg-Ising ladder with
the $XXZ$- and $XYZ$-Heisenberg intra-rung interaction can be found in the phases arising from the uniform configuration
with all $\tilde{s}_{1,i}^z=\frac{1}{2}$. While in the former model with the $XXZ$ intra-rung interaction, all ground-state phases are classical in their character \cite{verkholyak2012}, the emergent ground-state phases of the latter model with the more anisotropic $XYZ$ intra-rung coupling display significant quantum features. One may indeed identify
the following three quantum ground states for a particular case $e_0^+<e_0^-$ with all $\tilde{s}_{1,i}^z=\frac{1}{2}$:
\begin{itemize}
\item {\bf Quantum paramagnetic (QPM2) state} for $\frac{1}{2}|J_1^x-J_1^y|>|J_2+J_3|$:
    the equivalent transverse Ising chain $H^+$ (\ref{ham-pm}) is in the gapped disordered state
    with no spontaneous magnetization $\langle \tilde{s}_i^x\rangle=0$
    and non-zero magnetization $\langle \tilde{s}_i^z\rangle\neq 0$
    induced by the effective transverse field.
    For the initial Heisenberg-Ising ladder, one consequently gets the ground state with the prevailing dimer state
    $|\phi_{1,-}^{i}\rangle$ on the rungs.
 \item {\bf Stripe Rung (SR) state} for $\frac{1}{2}|J_1^x-J_1^y|<J_2+J_3$:
    the equivalent transverse Ising chain exhibits a spontaneous antiferromagnetic ordering
    with $\langle \tilde{s}_i^x\rangle=(-1)^i m_x\neq 0$.
    Due to relationships (\ref{spin-transf}), one obtains for the Heisenberg-Ising ladder
    $\langle s_{1,i}^z\rangle=-\langle s_{1,i+1}^z\rangle
     =\langle s_{2,i}^z\rangle=-\langle s_{2,i+1}^z\rangle\neq 0$.
    Thus, the Heisenberg-Ising ladder shows an antiferromagnetic order along the legs
    and ferromagnetic order along the rungs.
    The staggered magnetization as the relevant
    order parameter in this phase is non-zero and it exhibits
    evident quantum reduction of the magnetization given by:
    \[
    m_{\mathrm{SR}}^z=\frac{1}{2N}\sum_{i=1}^N\left(-1\right)^i\left(\langle s_{1,i}^z\rangle+\langle s_{2,i}^z\rangle\right)
     =\frac{1}{2}\left[1-\frac{\left(J_1^x-J_1^y\right)^2}{4\left(J_2+J_3\right)^2}\right]^{\frac{1}{8}}.
     \]
 \item {\bf Ferromagnetic (FM) state} for $\frac{1}{2}|J_1^x-J_1^y|<-(J_2+J_3)$:
    the effective transverse Ising chain shows a ferromagnetic ordering.
    Due to transformation relationships (\ref{spin-transf}), all the spins of the ladder have
    the same magnetization in the $z$-direction. The ground state corresponds to the ferromagnetic spin state
    \[
    m_{\mathrm{FM}}^z=\frac{1}{2N}\sum_{i=1}^N\left(\langle s_{1,i}^z\rangle+\langle s_{2,i}^z\rangle\right)
     =\frac{1}{2}\left[1-\frac{\left(J_1^x-J_1^y\right)^2}{4\left(J_2+J_3\right)^2}\right]^{\frac{1}{8}}.
     \]
\end{itemize}
\begin{figure}[ht]
\centerline{
\includegraphics[width=0.41\textwidth]{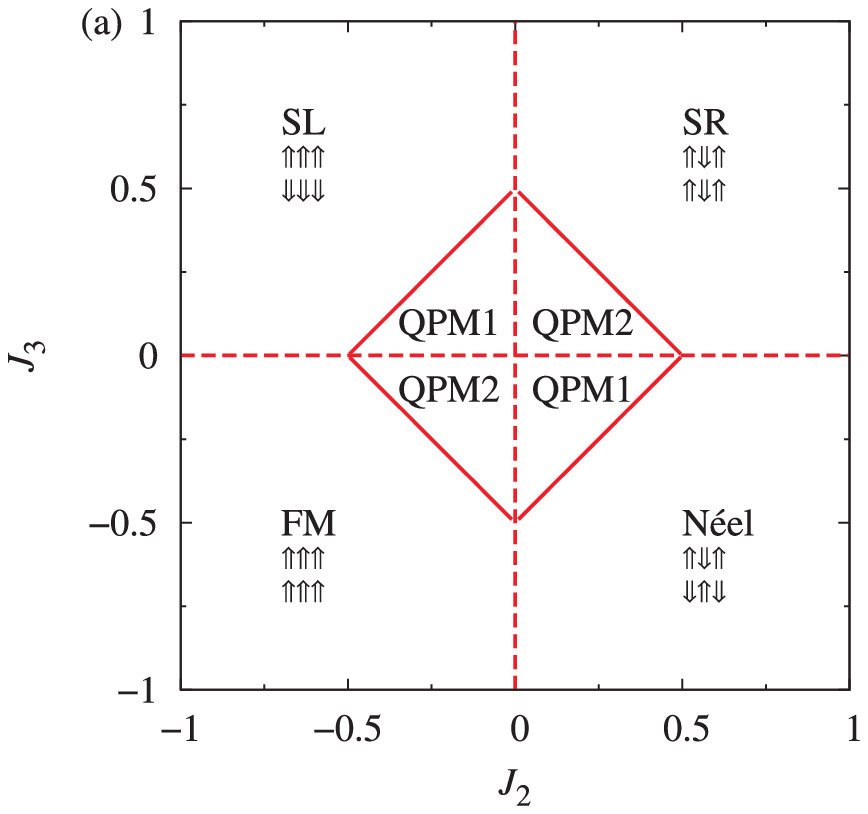}  
\hspace{5mm}
\includegraphics[width=0.41\textwidth]{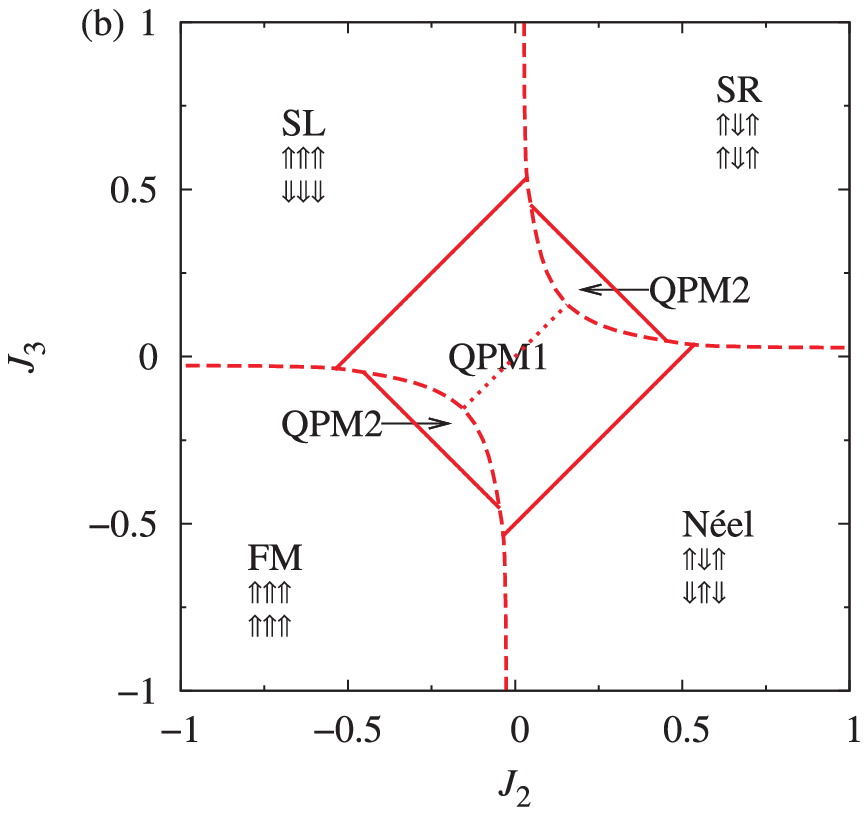} 
}
\centerline{
\includegraphics[width=0.41\textwidth]{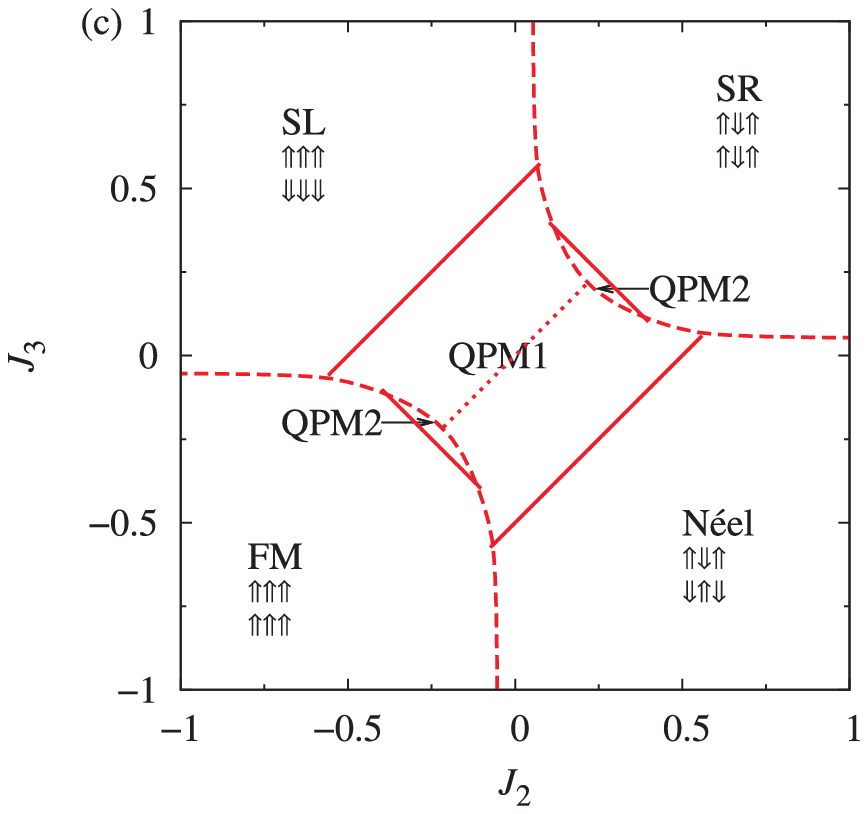}  
\hspace{5mm}
\includegraphics[width=0.41\textwidth]{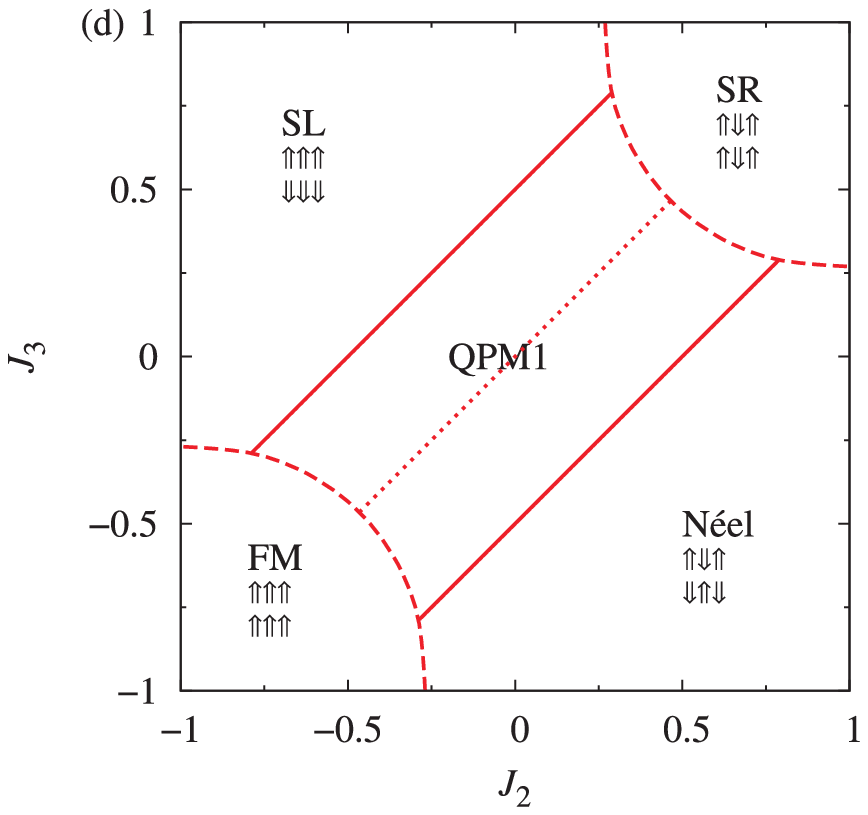}  
}
\caption{(Color online) Ground-state phase diagram of the $XZ$-Ising ladder
for $J_1^x=1$, $J_1^y=0$ and four different values of $J_1^z$:
(a) $J_1^z=0$; (b) $J_1^z=0.05$; (c) $J_1^z=0.1$; (d) $J_1^z=0.5$.
QPM1 is the paramagnetic phase with the prevailing rung states $|\phi_{0,0}^{i}\rangle$.
QPM2 is the paramagnetic phase with the prevailing rung states $|\phi_{1,-}^{i}\rangle$.
Dotted lines indicate the rung singlet-dimer state.
}
\label{fig-z-xz}
\end{figure}
\begin{figure}[!h]
\vspace{5mm}
\centerline{
\includegraphics[width=0.41\textwidth]{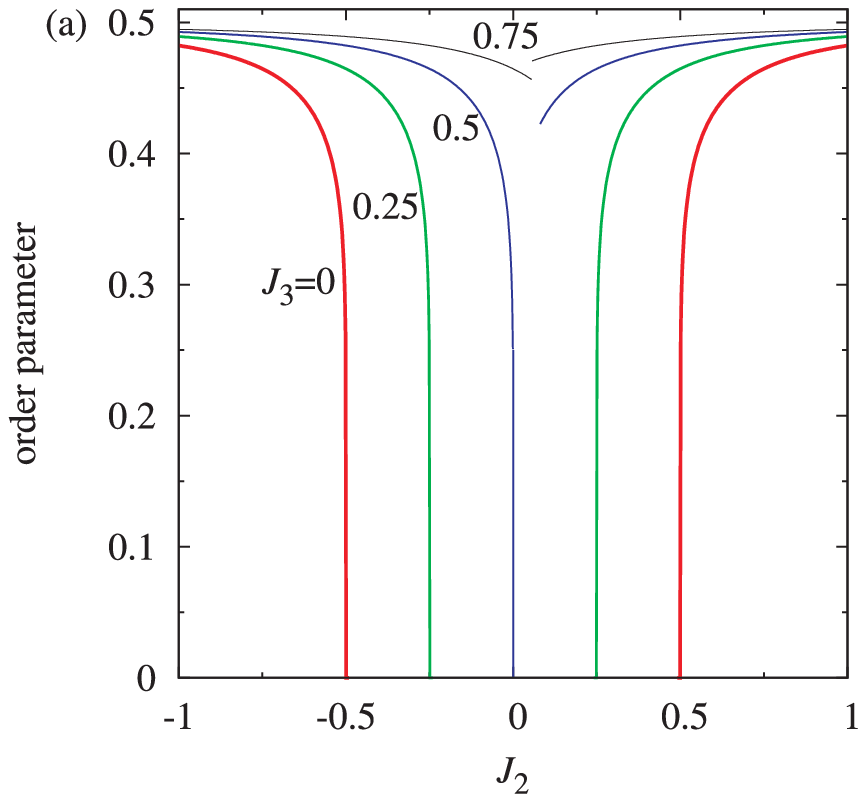} 
\hspace{5mm}
\includegraphics[width=0.41\textwidth]{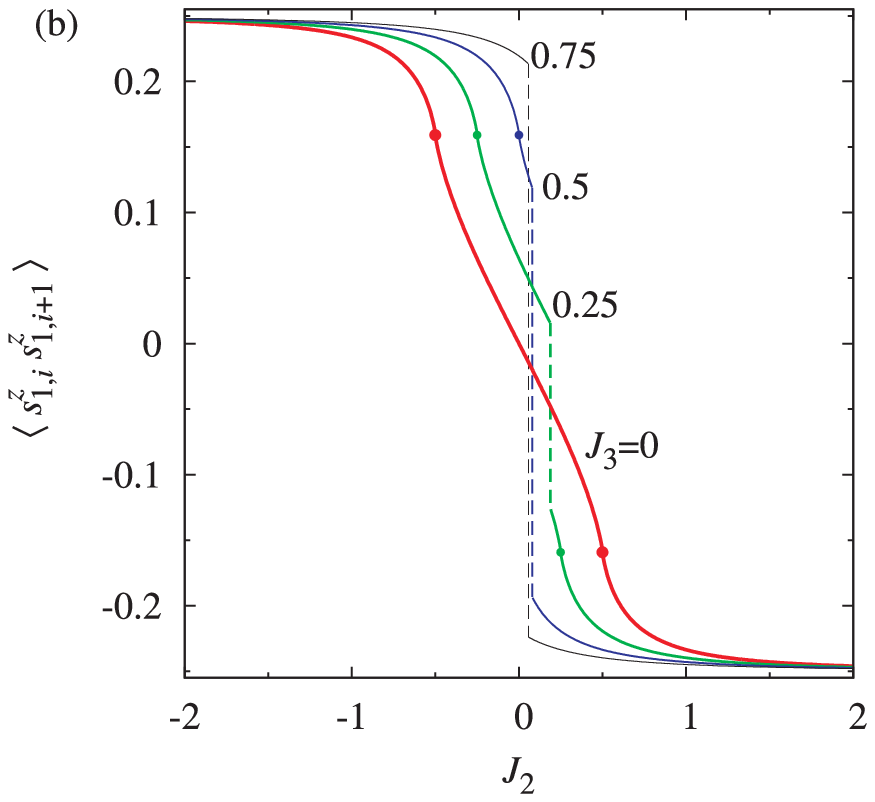} 
}
\caption{(Color online) Order parameters (a) and nearest-neighbor correlation function (b)
as a function of intra-leg interaction $J_2$ for $J_1^x=1$, $J_1^z=0.1$
and different $J_3=0,0.25,0.5,0.75$.
(a) The curves on the left correspond to $m_{\mathrm{SL}}$,
the curves on the right for $J_3=0.25,0.5,0.75$ ($J_3=0$) correspond to $m_{\mathrm{SR}}$ ($m_{\mathrm{AF}}$).
}
\label{fig-order}
\end{figure}
Altogether, it could be concluded that the $XY$ anisotropy in the Heisenberg intra-rung coupling is responsible for quantum features of otherwise classical SR and FM states and, moreover, it may also lead to the appearance of a new disordered phase QPM2.
Two paramagnetic phases QPM1 and QPM2 have quite similar features: they are both disordered states with the energy gap in their excitation spectrum and, consequently, their pair spin-spin correlation functions decay exponentially. Both quantum paramagnetic phases can be distinguished by the square of $z$th component of the total spin $(S_i^z)^2$ on $i$th rung, which is equal to $(S_i^z)^2=0(1)$ in QPM1 (QPM2).

Now, let us pay our attention to the ground-state phase diagram established for the particular case of $XZ$-Ising ladder as depicted
in figure~\ref{fig-z-xz} by considering $J_1^y = 0$. Assuming the $XZ$ intra-rung interaction,
one gets a striking competition  between the $X$--$X$ interaction along the rungs and the $Z$--$Z$ interaction along the legs,
while the additional  $Z$--$Z$ interaction along rungs acts generally against the $X$--$X$ interaction.
It should be also mentioned that one may recover some known examples from the ground-state phase diagram
of the $XZ$-Ising ladder presented in figure~\ref{fig-z-xz}. In fact, figure~\ref{fig-z-xz}~(a) shows the particular limiting case
of a quantum compass ladder with an additional diagonal frustrating Ising interaction. Let us follow the known results
of a simple quantum compass ladder \cite{brze2009} to be obtained from our model by disregarding the frustrating
Ising interaction $J_3=0$. Both Hamiltonians $H^+$ and $H^-$ (\ref{ham-pm}) become identical under this special condition
and, consequently, the ground state of the model is always two-fold degenerate due to the equality $e_0^-=e_0^+$.
The investigated model is either in the SL or FM state for $\frac{1}{2}J_1^x<-J_2$, either in QPM1 or QPM2 state for $\frac{1}{2}J_1^x>|J_2|$, either in SR or  N\'{e}el state for $\frac{1}{2}J_1^x<J_2$. Note that the quantum phase transition from the disordered to the long-range ordered state takes place at $|J_2|= \frac{1}{2}J_1^x$.

It is quite evident that the diagonal interaction $J_3$ removes the two-fold degeneracy of the ground state [see figure~\ref{fig-z-xz}~(a)]
by changing the effective spin interaction in the effective Hamiltonians $H^+$ and $H^-$. The effect of the additional $Z$--$Z$ intra-rung
interaction is shown in figure~\ref{fig-z-xz}~(b)--(d), where the results for a different relative strength of $J_1^z$ are presented.
It could be understood from (\ref{ham-pm}) that $J_1^z$ lowers initially the energy of the $\tilde{s}_{1,i}^z=-\frac{1}{2}$ configuration
by the amount $\frac{1}{2}NJ_1^z$ with respect to the energy of the $\tilde{s}_{1,i}^z=\frac{1}{2}$ configuration. Therefore, the regions of the phases, which correspond to the $\tilde{s}_{1,i}^z=\frac{1}{2}$ configuration, become smaller with increasing $J_1^z$. For sufficiently strong $J_1^z$, QPM2 phase can completely disappear from the ground-state phase diagram. Another distinctive feature is that $J_1^z$ makes the singlet-dimer phase favorable along the line $J_2=J_3$ inside the region of QPM1 state.

The relevant ground-state behavior can be supported by the dependencies of respective order parameters as displayed in figure~\ref{fig-order}~(a).
If there is no frustrating diagonal interaction, the model may stay in the disordered QPM1
or the ordered SL and N\'{e}el phases. The corresponding nearest-neighbor correlation function along the legs
shows a continuous change with a weak singularity at the quantum critical points indicated by filled circles
in figure~\ref{fig-order}~(b). The curve for  another particular case $J_3=0.25$ looks similar except that
the diagonal interaction of this strength leads to a direct phase transition between two disordered
quantum paramagnetic states QPM1 and QPM2. This unusual transition can be recognized from the relevant dependence
of the nearest-neighbor correlation function, which sustains a jump at this special critical point.
It is interesting to note that the further increase of a frustrating Ising interaction $J_3$ demolishes
both disordered phases QPM1 and QPM2. Thus, one may also detect the quantum phase transition between
two ordered SL and N\'{e}el phases, whereas the order parameters do not reach zero continuously in this particular case.

\begin{figure}[h]
\centerline{
\includegraphics[width=0.41\textwidth]{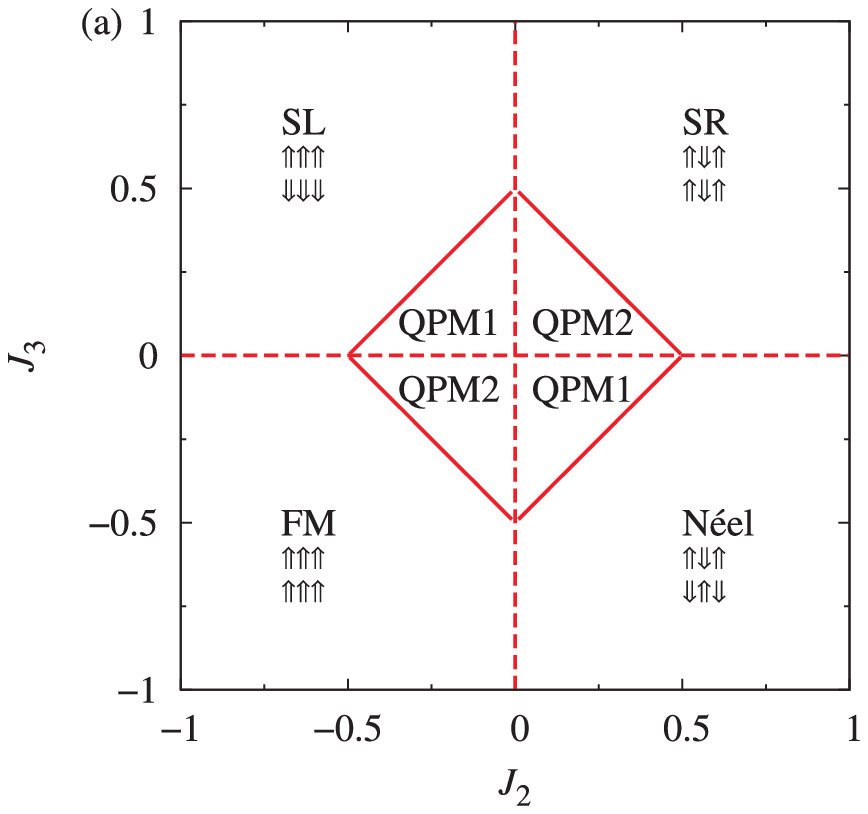}
\hspace{5mm}
\includegraphics[width=0.41\textwidth]{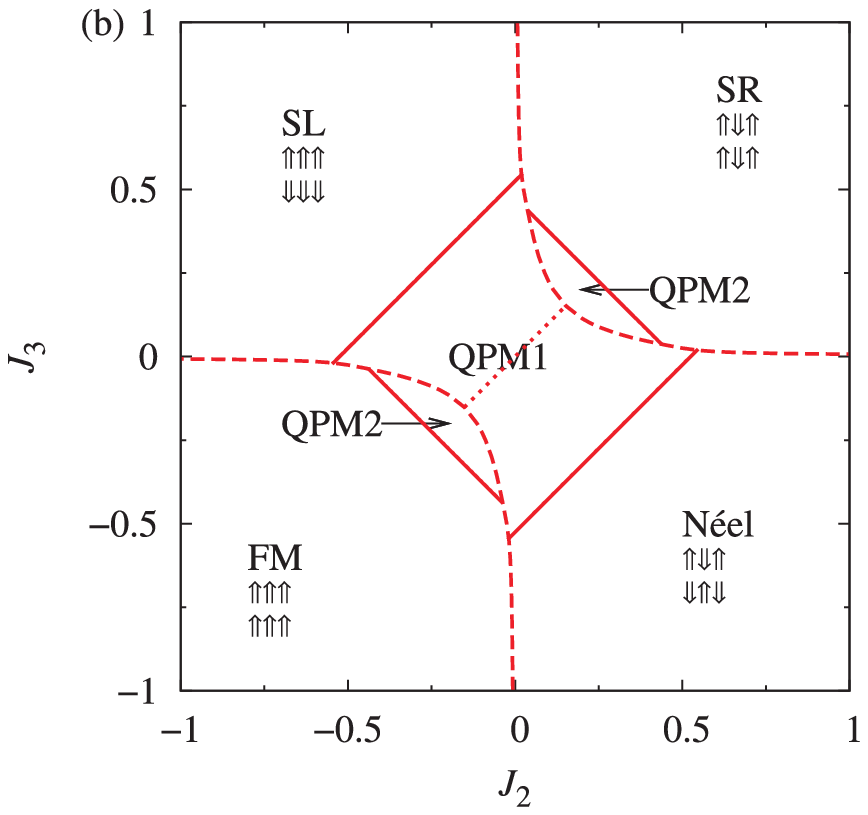}
}
\centerline{
\includegraphics[width=0.41\textwidth]{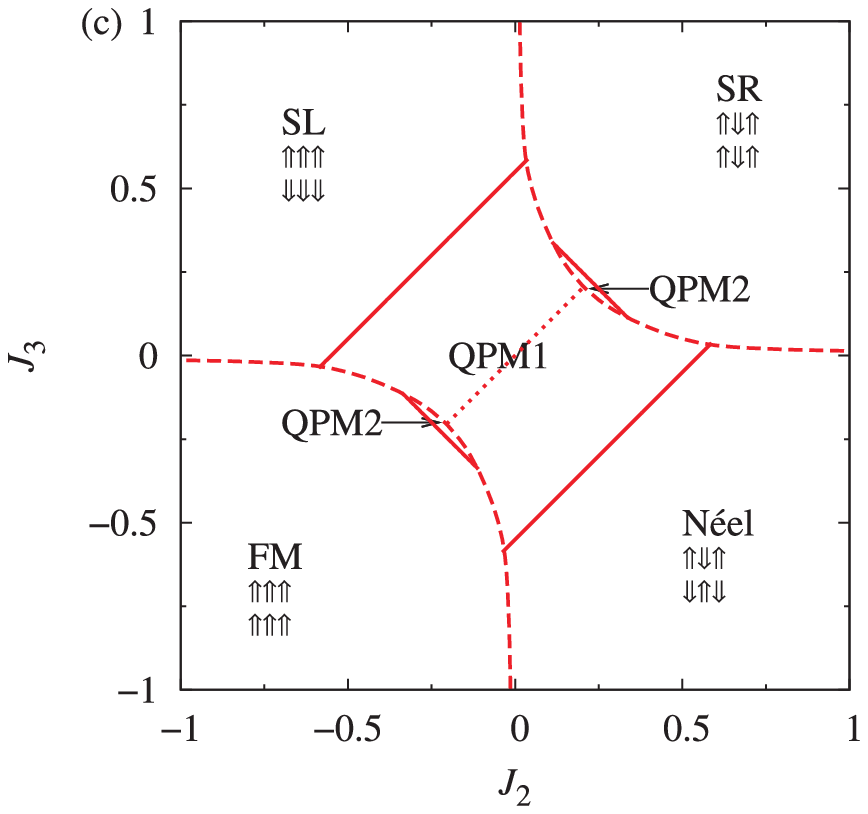}
\hspace{5mm}
\includegraphics[width=0.41\textwidth]{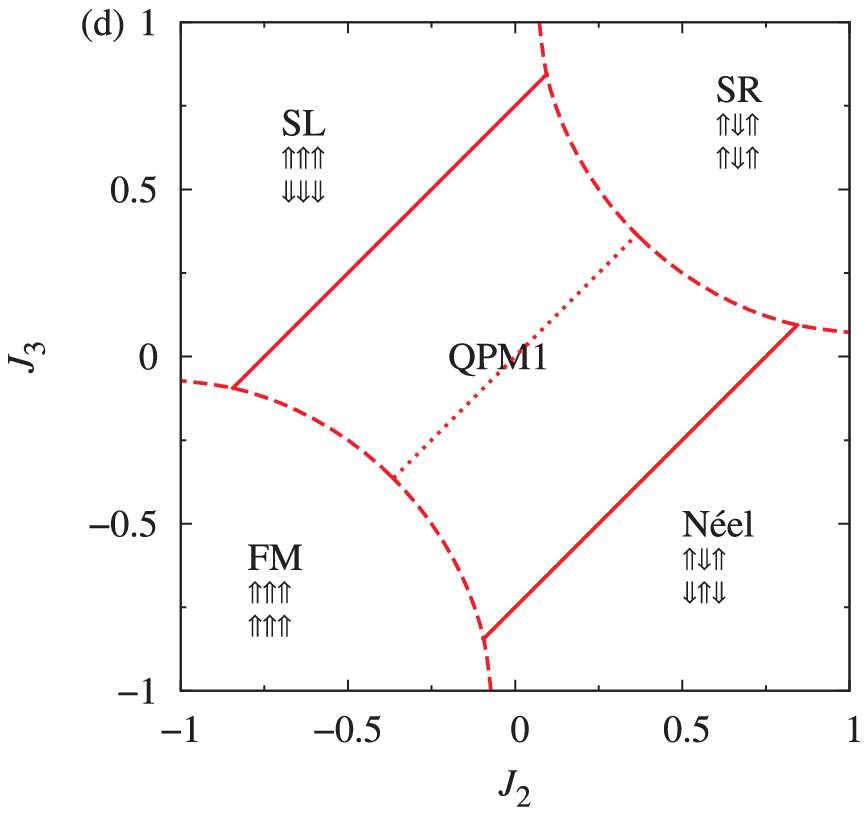}
}
\caption{(Color online) Ground-state phase diagram of the $XY$-Ising model
for $J_1^x=1$, $J_1^z=0$ and four different values of $J_1^y$:
(a) $J_1^y=0$; (b) $J_1^y=0.05$; (c) $J_1^y=0.1$; (d) $J_1^y=0.5$.
Dotted lines indicate the rung singlet-dimer state.
}
\label{fig-z-xy}
\end{figure}
\looseness=-1In figure~\ref{fig-z-xy}, the ground-state phase diagram of the $XY$-Ising ladder is depicted by considering
another particular case with $J_1^z=0$.
The effect of the $Y$--$Y$ intra-rung interaction has some similarities with the one of $Z$--$Z$ intra-rung interaction,
although the origin is completely different. The ground-state energy of $H^-$ is generally lowered with respect
to that of $H^+$, because $J_1^y>0$ increases the effective transverse field in $H^-$ and decreases it in $H^+$.
Owing to this fact, the regions of SL, QPM1, and N\'{e}el states that correspond to
the uniform configuration with all $\tilde{s}_{1,i}^z=-\frac{1}{2}$ get extended with an increase of $J_1^y$.
One may also generally conclude that $J_1^y>0$ acts against the QPM2 phase, which gradually shrinks with increasing
$J_1^y$ until it completely disappears at $J_1^y = \frac{1}{2} J_1^x$.
Similarly to the case with the $Z$--$Z$ intra-rung interaction, the $Y$--$Y$ intra-rung interaction also induces
the presence of the rung singlet-dimer state along a special line $J_2=J_3$.

\section{Conclusions}
In the present paper, the effect of the most general $XYZ$ anisotropy in the intra-rung interaction
on the ground state of the spin-$\frac{1}{2}$ Heisenberg-Ising two-leg ladder was investigated in detail.
It has been shown that the most general kind of anisotropy, which breaks the rotational symmetry of the Heisenberg interaction,
may lead to the appearance of new quantum phases in the ground-state phase diagram. We have also considered
the special case of quantum compass ladder with an additional frustrated diagonal interaction
and showed that the singlet-dimer phase cannot appear in this particular case. The order parameters and
the nearest-neighbor correlation function were calculated and  analyzed in detail in the ground state.
It has been demonstrated that the relevant behavior of the correlation function can help us to reveal
the quantum phase transition between two different disordered quantum paramagnetic phases.

\section*{Acknowledgements}
J.S. acknowledges the financial support provided by ERDF EU (European Union European regional development fund) grant
under the contract ITMS 26220120005 (activity 3.2).



\vspace{-10mm}
\ukrainianpart

\title{Основний стан спін-1/2 двоногої драбинки Гайзенберґа-Ізинґа з $XYZ$ взаємодією
вздовж щаблів}

\author{Т.~Верхоляк\refaddr{label1}, Й.~Стречка\refaddr{label2}}

\addresses{
\addr{label1} Інститут фізики конденсованих систем НАН України,
вул. Свєнціцького, 1, 79011 Львів, Україна
\addr{label2} Кафедра теоретичної фізики і астрофізики, Університет П.Й. Шафарика,
парк Ангелінум, 9, 04001 Кошиці, Словацька республіка
}

\makeukrtitle

\begin{abstract}
\tolerance=3000%
Квантову спін-1/2 двоногу драбинку з анізотропною $XYZ$ взаємодією Гайзенберґа
вздовж щаблів і Ізин\-ґо\-вою взаємодією між спінами на сусідніх щаблях
розглянуто в межах строгого підходу, який ґрунтується на унітарному перетворенні.
Частковий випадок моделі з $X$--$X$ взаємодією вздовж щаблів відображає
модель квантового компасу на драбинці з додатковими діагональними Ізинґовими взаємодіями.
Використовючи унітарне перетворення, модель можна звести до поперечного ланцюжка Ізинґа
з ком\-по\-зит\-ни\-ми спінами, і як наслідок основний стан можна знайти строго.
Ми отримуємо фазову діаграму основного стану і аналізуємо взаємовплив конкуренції кількох факторів:
$XYZ$ анізотропії взаємодії Гайзенберґа, Ізинґової взаємодії вздовж ланцюжків та
фрустрованої діагональної Ізинґової взаємодії.
Розглянута модель демонструє складні фазові діаграми основного стану,
включаючи декілька нетипових квантових впорядкованих станів,
два різних невпорядкованих квантових станів, а також квантові фазові переходи
першого і другого порядку між цими фазами.
\keywords квантова спін-1/2 драбинка, точні результати, основний стан, квантовий фазовий перехід

\end{abstract}

\end{document}